\documentclass[twocolumn,aps,prd,amsfonts,amsmath,amssymb,nofootinbib,longbibliography,preprintnumbers]{revtex4-1}

\usepackage{ifpdf}

\ifpdf
    \usepackage[pdftex,colorlinks=true,plainpages=false,pdfpagelabels,breaklinks=true]{hyperref}
    \usepackage[pdftex]{color}
\else
    \usepackage[colorlinks=true]{hyperref}
\fi

\newcommand{\half}{\frac{1}{2}}
\newcommand{\bbar}[1]{\overline{#1}}

\begin{document}

\preprint{IPMU14-0037}
\title{Emergent Lorentz Signature, Fermions, and the Standard Model}

\author{John Kehayias}
 \email{john.kehayias@ipmu.jp}
\author{Shinji Mukohyama}
 \email{shinji.mukohyama@ipmu.jp}
 \affiliation{Kavli Institute for the Physics and Mathematics of the Universe (WPI)\\
  Todai Institutes for Advanced Study, The University of Tokyo\\
  Kashiwa, Chiba  277-8582, Japan}

\author{Jean-Philippe Uzan}
 \email{uzan@iap.fr}
 \affiliation{Institut d'Astrophysique de Paris, Universit\'e Pierre \& Marie Curie\\
  CNRS-UMR 7095, 98 bis, Bd Arago, 75014 Paris, France}
 \affiliation{Sorbonne Universit\'es, Institut Lagrange de Paris,\\
  98 bis bd Arago, 75014 Paris (France)}

\begin{abstract}

  \noindent
  This article investigates the construction of fermions and the
  formulation of the Standard Model of particle physics in a theory in
  which the Lorentz signature emerges from an underlying microscopic
  purely Euclidean $SO(4)$ theory. Couplings to a clock field are
  responsible for triggering the change of signature of the effective
  metric in which the standard fields propagate. We demonstrate that
  Weyl and Majorana fermions can be constructed in this
  framework. This construction differs from other studies of Euclidean
  fermions, as the coupling to the clock field allows us to write down
  an action which flows to the usual action in Minkowski spacetime. We
  then show how the Standard Model can be obtained in this theory and
  consider the constraints on non-Standard Model operators which can
  appear in the QED sector due to CPT and Lorentz violation.

\end{abstract}

\maketitle

\section{Introduction}

Part of the art of theoretical physics is to find the mathematical
structures that allow us to formalize and simplify the laws of
nature. These structures include the description of spacetime
(dimension, topology, \ldots) and matter and their interactions
(fields, symmetries, \ldots). While there is a large amount of freedom
in the choice of these mathematical structures, the developments of
theoretical physics have taught us that some of them are better suited
to describe certain classes of phenomena. However, these choices are
only validated by the mathematical consistency of the theory and, in
the end, by the agreement of their predictions with experiments.

Among all of these structures, and in the framework of metric theories
of gravitation, the signature of the metric is in principle
arbitrary. It seems that on the scales that have been probed so far
there is the need for only one time dimension and three spatial
dimensions. It is also now universally accepted that the relativistic
structure is a central ingredient of the construction of any realistic
field theory, in particular as the cleanest way to implement the
notion of causality. Spacetime enjoys a locally Minkowski structure
and, when gravity is included, the equivalence principle implies (this
is not a theoretical requirement, but an experimental fact, required
at a given accuracy) that all the fields are universally coupled to
the same Lorentzian metric. Thus, we usually take for granted that
spacetime is 4-dimensional manifold endowed with a metric of mixed
signature, e.g.~$(-,+,+,+)$.

While the existence of two time directions may lead to confusion
\cite{V4,V42}, several models for the birth of the universe
\cite{nothing,nothing2,nothing3,nothing4} are based on a change of
signature via an instanton in which a Riemannian and a Lorentzian
manifold are joined across a hypersurface. While there is no time in
the Euclidean region, with signature $(+,+,+,+)$, it flips to
$(-,+,+,+)$ across this hypersurface, which may be thought of as the
origin of time from the Lorentzian point of view. Eddington even
suggested \cite{G85} that it can flip across some surface to
$(-,-,+,+)$ and signature flips also arise in brane or loop quantum
cosmology \cite{lqc,lqc2,lqc3}.

It is legitimate to investigate whether the signature of the metric is
only a convenient way to implement causality or whether it is just a
property of an effective description of a microscopic theory in which
there is no such notion. In
Ref.~\cite{Mukohyama:2013ew,*emergentessay}, two of us have proposed
that at the microscopic level the metric is Riemannian and that the
Lorentzian structure, usually thought of as fundamental, is in fact an
effective property that emerges in some regions of a 4-dimensional
space with a positive definite metric. There has been some related
work in the past --- for instance, the work by Barbero \cite{BarberoG.:1995ud}
(with more than second-order derivatives in the equations of motion,
however), or in Einstein-Aether theory
\cite{BarberoG.:2003qm,*Foster:2005ec} (although without an order
parameter connecting the Euclidean and Lorentzian theories) and scalar
gravity \cite{Girelli:2008qp}. We argued that a decent classical field
theory for scalars, vectors, and spinors in flat spacetime can be
constructed, and that gravity can be included under the form of a
covariant Galileon theory instead of general relativity. This
mechanism of emergent Lorentz signature may also serve as a new way to
circumvent the issue of non-unitarity in some higher-derivative
quantum gravity theories \cite{Mukohyama:2013gra, Muneyuki:2013aba}.

Among the gaps emphasized in this work, we have pointed out that (1)
the construction is restricted to classical field theory and the
spinor sector suffers from a severe fine-tuning to ensure CPT
invariance (see e.g.~Ref.~\cite{Toma:2012xa} and references therein
for recent constraints on CPT violation), and that (2) it requires the
construction of Majorana and Weyl spinors in order to formulate the
Standard Model (SM) and its extensions.

It is well known that Majorana fermions are technically impossible to
construct in a 4d Euclidean theory, but several authors have found
alternative constructions \cite{Mehta:1986mi, Mehta:1991ve,
  vanNieuwenhuizen:1996tv, vanNieuwenhuizen:1996ip,
  Wetterich:2010ni}. However, these techniques are often aimed at a
Wick rotation to or from a Lorentzian theory and can involve doubling
the fermion degrees of freedom or other aspects which are ill suited
to our application. With the aim of developing a theory which flows to
the usual actions in Minkowski space (which may look very different in
Euclidean space) and the available couplings to the clock field, we
arrive at a new formulation for Weyl and Majorana spinors. As our
goals and setting are different than in previous studies, we do not
need to use the techniques employed there, such as fermion doubling or
the ad hoc construction of different spinors. The Weyl spinors and
coupling to the clock field allow us to directly construct an emergent
version of the SM, with its chiral and metric structure inherited from
an originally Euclidean theory.

This article is organized as follows. In Section
\ref{sec:emerg-lorentz-sign} we briefly review the construction given
in Ref.~\cite{Mukohyama:2013ew,*emergentessay}. Following that, in
Section \ref{sec:weyl-major-ferm} we extend the fermion sector to
include Weyl and Majorana fermions, which is quite distinct from the
usual considerations in Euclidean space. For fermions, an alternative
``derivation'' of several of the choices in this construction are
detailed in the Appendix. In Section \ref{sec:standard-model} we then
show how to construct the Standard Model in this framework of an
emergent Lorentzian metric. There are additional operators which can
arise in this theory and, in Section \ref{sec:constraints-qed}, we
categorize and analyze the constraints on such operators in the QED
sector of the SM. Finally, we gather further comments, conclusions,
and future directions in Section \ref{sec:disc-concl-outl}.

\section{Emergent Lorentz Signature}
\label{sec:emerg-lorentz-sign}

In this section we briefly lay out our conventions and review the
construction given in Ref.~\cite{Mukohyama:2013ew,*emergentessay} for a theory with
an effective Lorentz signature emerging from a locally Euclidean
metric. The Minkowski metric, $\eta_{\mu\nu}$, is mostly positive with
signature $(-, +, +, +)$, while the Euclidean metric has positive
signature and is denoted $\delta_{\mu\nu}$.

\subsection{Basics of the mechanism}
From the point of view of the Euclidean theory, at the fundamental
level, there is no concept of time (one cannot single out a privileged
direction) until the clock field, $\phi$, picks out a direction
through its derivative having a nonzero vacuum expectation value
(vev). We will always work in some patch $\mathcal{M}_0$ where this
vev can be considered a constant,
\begin{equation}
  \label{eq:clockvev}
  \partial_\mu\phi = M^2n_\mu,
\end{equation}
with $M$ a mass scale for units and $n_\mu$ a constant unit vector which
now defines a particular direction, related to the notion of time (the
direction which will change signature in the effective metric). Thus,
we have
\begin{equation}
  \label{eq:dt}
  \mathrm{d}t = n_\mu\mathrm{d}x^\mu
\end{equation}
and choose
\begin{equation}
  \label{eq:t}
  t \equiv \frac{\phi}{M^2}.
\end{equation}
The other three coordinates (with positive signature) are the
coordinates of a hypersurface normal to $n_\mu$.

We can now write down actions in the Euclidean theory, which will flow
to a Minkowski theory (in the sense that the fields propagate in an
effective Minkowski metric) when restricted to $\mathcal{M}_0$ after the
gradient of the clock field has a vev. Here we just summarize the
results obtained in Ref.~\cite{Mukohyama:2013ew,*emergentessay}, which
has further details. 

For a scalar field $\chi$ with potential $V(\chi)$, we consider a Euclidean
action of the form
\begin{align}
  \label{eq:scalar_e}
  S_\chi = \int\mathrm{d}^4x\left[ -\half\delta^{\mu\nu}\partial_\mu\chi\partial_\nu\chi - V\left(\chi\right) \right.\nonumber\\
\left. + \frac{1}{M^4}\left( \delta^{\mu\nu}\partial_\mu\phi\partial_\nu\chi \right)^2 \right].
\end{align}
In $\mathcal{M}_0$ the last term becomes $M^4(\partial_t\chi)^2$ so that the
above action leads to the usual action for a scalar field in Minkowski
space,
\begin{equation}
  \label{eq:scalar_m}
  S_\chi = \int\mathrm{d}t\mathrm{d}^3x \left[ -\half\eta^{\mu\nu}\partial_\mu\chi\partial_\nu\chi - V\left(\chi\right) \right].
\end{equation}

The case of a vector field, $A_\mu$, with field strength tensor
$F_{\mu\nu}^E$ (the $E$ denotes that indices are raised/lowered with
the Euclidean metric) is also straightforward. The action
\begin{equation}
  \label{eq:vector_e}
  S_A = \frac{1}{4}\int\mathrm{d}^4x\left[ -F_{\mu\nu}^EF^{\mu\nu}_E + \frac{4}{M^4}F^{\mu\rho}_EF^\nu_{E\rho}\partial_\mu\phi\partial_\nu\phi \right],
\end{equation}
with the second term equaling $4\delta^{ij}F_{0i}F_{0j}$ in
$\mathcal{M}_0$, becomes the standard Maxwell action for a vector
field in Minkowski spacetime,
\begin{equation}
  \label{eq:vector_m}
  S_A = -\frac{1}{4}\int\mathrm{d}t\mathrm{d}^3x\eta^{\mu\alpha}\eta^{\nu\beta}F_{\alpha\beta}F_{\mu\nu}.
\end{equation}

\subsection{Dirac Fermions}
\label{sec:diracferm}
We will now consider Dirac fermions, which will require a bit more
detail and care, as we have to be careful with the Clifford algebra
and the gamma matrices to build a proper action. This will be extended
to Weyl and Majorana fermions in the following section, while a more
(Clifford) basis agnostic derivation of these conventions can be found
in the Appendix.

In general, the gamma matrices $\gamma^\mu$ satisfy\footnote{The overall sign
  can be changed by a factor of $i$ in the gamma matrices, changing
  the Hermiticity of the matrices.}
\begin{equation}
  \label{eq:ggmetric}
  \left\{\gamma^\mu, \gamma^\nu\right\} = -2g^{\mu\nu},
\end{equation}
for a metric $g^{\mu\nu}$. These matrices generate the group ($SO(4)$ or
$SO(3,1)$ in our case) generators
\begin{equation}
  \label{eq:gensgen}
  S^{\mu\nu} \equiv \frac{i}{4}[\gamma^\mu, \gamma^\nu].
\end{equation}

In Minkowski space we will use the common Weyl or chiral
representation with the Pauli matrices $\sigma^\mu \equiv (1,
\sigma^i), \bar{\sigma}^\mu \equiv (1, -\sigma^i)$, and the gamma matrices
\begin{equation}
  \label{eq:gamma_m}
  \gamma^\mu_M \equiv \left(
    \begin{matrix}
      0 & \sigma^\mu\\
      \bar{\sigma}^\mu & 0
    \end{matrix}
  \right).
\end{equation}
We define\footnote{Note that this definition includes a minus sign.}
$\gamma^5_M$ as
\begin{equation}
  \label{eq:g5_m}
  \gamma^5_M \equiv -i\gamma^0_M \gamma^1_M \gamma^2_M \gamma^3_M = \mathrm{diag} (1,1,-1,-1),
\end{equation}
which is Hermitian, squares to $1$, and anticommutes with all $\gamma^\mu_M$.

A $4$-component Dirac spinor, $\psi_M$, transforms as
\begin{equation}
  \label{eq:psim}
  \psi_M \rightarrow \Lambda_{M,\half}\psi_M, \quad \Lambda_{M,\half} = \exp\left[-\frac{i}{2}\omega_{\mu\nu}S^{\mu\nu}_M\right]
\end{equation}
with $\omega$ an antisymmetric tensor and $\Lambda_{M,\half}$ not unitary in
general. In order to form Lorentz invariants for an action, we define
the usual barred spinor,
\begin{equation}
  \label{eq:psibarm}
  \bar{\psi}_M \equiv \psi_M^\dagger\gamma^0_M, \quad \bar{\psi}_M \rightarrow \bar{\psi}_M\Lambda^{-1}_{M,\half}.
\end{equation}
The standard action for the Dirac field in Minkowski space is given by
\begin{equation}
  \label{eq:diracsm}
  S^M_\psi = \int\mathrm{d}^4x~\bar{\psi}_M\left( \frac{i}{2}\gamma^\mu_M\overleftrightarrow{\partial}_\mu - m \right)\psi_M.
\end{equation}

In Euclidean space the gamma matrices are chosen as
\begin{equation}
  \label{eq:gammae}
  \gamma^0_E \equiv i\gamma^5_M, \quad \gamma^i_E \equiv \gamma^i_M,
\end{equation}
and $\gamma^5_E$ satisfies the same properties, now defined as
\begin{equation}
  \label{eq:g5e}
  \gamma^5_E \equiv \gamma^0_E \gamma^1_E \gamma^2_E \gamma^3_E = \gamma^0_M.
\end{equation}
The generators of $SO(4)$, $S^{\mu\nu}_E$, are now Hermitian and so
$\Lambda_{E,\half}$ is a unitary transformation of the $4$-component spinor
$\psi_E$,
\begin{equation}
  \label{eq:psie}
  \psi_E \rightarrow \Lambda_{E,\half}\psi_E, \quad \Lambda_{E,\half} = \exp\left[-\frac{i}{2}\omega_{\mu\nu}S^{\mu\nu}_E\right].
\end{equation}
Both $\bar{\psi}_E (=\psi^\dagger_E\gamma^0_M = \psi^\dagger\gamma^5_E)$ and
$\psi^\dagger$ transform the same way,
\begin{equation}
  \label{eq:barpsie}
  \bar{\psi}_E \rightarrow \bar{\psi}_E\Lambda^{-1}_{E,\half}, \quad \psi^\dagger_E \rightarrow \psi^\dagger_E\Lambda^{-1}_{E,\half},
\end{equation}
and can form $SO(4)$ invariants with $\psi_E$. We will favor the bar
notation to make the connection to the Lorentzian theory explicit.

The following Euclidean action,
\begin{align}
  \label{eq:dirace}
  S_\psi &= \int\mathrm{d}^4x\left\{\bar{\psi}_E\left( \frac{i}{2}\gamma^\mu_E\stackrel{\leftrightarrow}{\partial_\mu} - m \right)\psi_E\right. \\
      & \left. + \frac{1}{2M^2}\delta^{\mu\nu}\left[ \left(i\bar{\psi}\gamma^5_E\stackrel{\leftrightarrow}{\partial_\mu}\psi\right) -\left(i\bar{\psi}\gamma^\rho_E\stackrel{\leftrightarrow}{\partial_\mu}\psi\right)\partial_\rho \phi\right]\partial_\nu\phi\right\},\nonumber
\end{align}
becomes the Minkowski Dirac action, Eq.~\eqref{eq:diracsm}, after the
clock field picks out a direction in $\mathcal{M}_0$.

\section{Weyl and Majorana Fermions}
\label{sec:weyl-major-ferm}

We now extend the above procedure for Weyl and Majorana fermions. By
a Weyl fermion, we mean a 4-component spinor that is an eigenstate of
$\gamma^5$,
\begin{equation}
  \label{eq:weyldefn}
  \gamma^5_{E,M}\psi_\pm^{E,M} = \pm\psi_\pm^{E,M},
\end{equation}
and we recall that in the representation used above, $\gamma^5_E \neq \gamma^5_M$.

It is important to note that the $\gamma^\mu_E$ representation we have used
is \emph{not} the same as the Weyl or chiral representation: it does
not make manifest the algebra isomorphism\footnote{It is important to
  note that unlike in the Lorentzian case, the representations of
  these $SU(2)$s are \emph{not} related by complex conjugation. In
  other words, a 2-component spinor and its complex conjugate
  transform under the same $SU(2)$ and can make an $SO(4)$
  invariant. For a review of 2-component spinors, see
  Ref.~\cite{Dreiner:2008tw} and references therein, as well as
  Ref.~\cite{McKeon:2001pm} for Euclidean space.} $SO(4) = SU(2)_-
\times SU(2)_+$. In other words, a general $SO(4)$ transformation of a
$4$-component spinor, $\psi^E$, in this description does not separate
into two $2$-component spinors (the top/bottom of the $4$-component
spinor) transforming in separate $SU(2)$s. This is also why we have
suppressed all spinor indices, as there is not the usual separation
into dotted and undotted indices labeling the different $SU(2)$s.

However, the eigenstates of $\gamma^5_E$ take the following form,
\begin{equation}
  \label{eq:2compe}
  \psi^E_\pm = \left(
    \begin{matrix}
      \xi_\pm\\
      \pm\xi_\pm
    \end{matrix}
  \right),
\end{equation}
with $\xi_\pm$ transforming as a $2$-component spinor under
$SU(2)_\pm$. We can also form the usual projection matrices with $(1
\pm \gamma^5_E)/2$. For $\psi^E_\pm$ then, we can make a direct
connection with $2$-component spinors in this formalism. It should be
noted, however, that it is best to work in one form or the other, as
the decomposition between 4- and 2-component spinors is completely
different in our Euclidean and Lorentzian theories.\footnote{See
  Ref.~\cite{Dreiner:2008tw} and references therein for details in
  translating between 2- and 4-component spinors, in 4d Minkowski in
  particular.} In the Lorentzian theory, the Weyl spinors are of the
form
\begin{equation}
  \label{eq:2compwm}
  \psi^M_{L(-)} = \left(
    \begin{matrix}
      \xi_-\\
      0
    \end{matrix}
  \right), \qquad
  \psi^M_{R(+)} = \left(
    \begin{matrix}
      0\\
      \xi_+
    \end{matrix}
  \right).
\end{equation}

To construct an action in the Euclidean theory which will become the
appropriate action in the Lorentzian theory we might try using the
terms
\begin{equation}
  \label{eq:weyl0terms}
  i\bbar{\psi^E_\pm}\gamma^\mu_E\partial_\mu\psi^E_\pm~, \qquad \delta^{\mu\nu}\left(i\bbar{\psi^E_\pm}\gamma^\rho_E\partial_\mu\psi^E_\pm\right)\partial_\rho\phi\partial_\nu\phi,
\end{equation}
but unfortunately they vanish identically. Instead, we can construct an 
appropriate action as
\begin{align}
  \label{eq:weylm2}
  S_\pm = \frac{1}{4M^2}\int\mathrm{d}t\mathrm{d}^3x&\left[ \bbar{\psi}^E_\pm\gamma^5_E\left( i\delta^{\mu\nu} - \gamma^\nu_E\gamma^\mu_E \right)\partial_\mu\psi^E_\pm\partial_\nu\phi\right. \nonumber\\
  &\left. + ~\rm{h.c.}~ \vphantom{\bbar{\psi}^E_\pm}\right].
\end{align}
After the gradient of the clock field has a vev on $\mathcal{M}_0$,
this becomes the standard Lorentzian action for $2$-component Weyl
spinors,
\begin{equation}
  \label{eq:weyle2}
  S_\pm = \int\mathrm{d}t\mathrm{d}^3x
  \begin{cases}
    i\xi_{-,L}^\dag\bar{\sigma}^\mu\partial_\mu\xi_{-,L}\\
    i\xi_{+,R}^\dag\sigma^\mu\partial_\mu\xi_{+,R}
  \end{cases},
\end{equation}
where the subscript indicates the $SU(2)$ representation from the
Euclidean ($_\mp$) to Lorentzian ($_{L,R}$). To connect to the
$4$-component spinors, we recognize that, once the gradient of the
clock field has a vev, we want eigenstates of $\gamma^5_M$,
\begin{equation}
  \label{eq:weylmtoe}
  \gamma^5_E\psi^E_\pm = \pm\psi^E_\pm \quad \rightarrow \quad \gamma^5_M\psi^M_\pm = \pm\psi^M_\pm.
\end{equation}
This naturally comes out of the action of Eq.~\eqref{eq:weylm2}. By
inserting $(-i*i)$ in the second term and using the properties of the
gamma matrices the action becomes
\begin{equation}
  \label{eq:weylm4}
  S = \int\mathrm{d}t\mathrm{d}^3x~i\bbar{\psi}^M_\pm\gamma^\mu\partial_\mu\psi^M_\pm.
\end{equation}

Finally, we also want to incorporate Majorana spinors (representing
fermions which are their own antiparticles). As is well known, we
cannot directly have a Majorana spinor in $4$d Euclidean space:
$\psi^\mathcal{C}_E = \psi_E$ is only satisfied for the zero spinor,
where $\psi^\mathcal{C}_E \equiv C_E\bbar{\psi}^T$ is the Euclidean
charge conjugate spinor and $C_E$ will be defined below (see
Eq.~\eqref{eq:ce}).

However, using the above formulation of Weyl spinors, we can write
down a Lagrangian for a single Weyl fermion with a mass term. This
captures the physical properties of a Majorana spinor, and in the
Lorentzian theory this will correspond to the usual Majorana spinor (a
self-conjugate 4-spinor). From our form of Weyl spinors,
Eq.~\eqref{eq:2compe}, we write a Majorana mass term (the right-hand
side is exactly the Lorentzian 2-component form as we have rotated
$\psi$ to change the signs) as
\begin{equation}
  \label{eq:majme}
  \frac{1}{4} m(\psi^E_\pm)^TC_E\psi^E_\pm~+~\mathrm{h.c.} = \half m\xi_\pm\xi_\pm~+~\mathrm{h.c.}
\end{equation}
with $T$ denoting the transpose, and where we need the matrix $C_E$ to
have the term be $SO(4)$ invariant (i.e.~to provide the (suppressed)
invariant to combine the $\xi_\pm$ spinors as in the usual 2-component
formalism). In other words, we require
\begin{equation}
  \label{eq:ceprop}
  \Lambda^T_{E,\half}C_E = C_E \Lambda^{-1}_{E,\half},
\end{equation}
which is satisfied by the matrix\footnote{We are not using explicit
  spinor indices, so we consider this as a numerical identification.}
\begin{equation}
  \label{eq:ce}
  C_E = \gamma^1_E\gamma^3_E,
\end{equation}
with properties
\begin{equation}
  \label{eq:cmprop}
  C_E^T = C_E^\dag = C_E^{-1} = -C_E.
\end{equation}
This is similar to the numerical structure of a charge conjugation
matrix,\footnote{We have not shown how it operates directly on
  (anti)particles. Also, the matrix satisfies $C_E^{-1}\gamma^\mu_EC_E
  = (\gamma^\mu_E)^T$ rather than giving $-(\gamma^\mu_E)^T$ as in the
  usual Minkowski space definition.} but again, we cannot enforce that
a spinor is self-conjugate and nontrivial in the 4d Euclidean theory
(as one can see directly given $C_E$ above). When we move to the
Lorentzian theory, this matrix becomes
\begin{equation}
  \label{eq:tildecm}
  \widetilde{C}_M = \gamma^1_M\gamma^3_M,
\end{equation}
which is almost the Lorentzian charge conjugation matrix. If we use a
factor\footnote{The sign is automatic from the left-handed field, or
  through a field rotation for the right-handed field (the sign of the
  Majorana mass term can be changed freely).} of $-\gamma^5_M$ in the
mass term from the property of the (now Lorentzian) Weyl spinor, we
can now identify (again, as a numerical identity through direct
computation of the necessary properties) this with the Lorentzian
charge conjugation matrix $C_M$,
\begin{equation}
  \label{eq:cm}
  C_M = -\gamma^5_M\gamma^1_M\gamma^3_M.
\end{equation}
The structure of this mass term,
\begin{equation}
  \label{eq:majmm}
  \half m\psi^T_{\pm,M}C_M\psi_{\pm,M},
\end{equation}
is exactly a Majorana mass term with the identification of the
Majorana condition,
\begin{equation}
  \label{eq:majcondm}
  \psi^C_M \equiv C_M\bbar{\psi}^T_M = \psi_M ~~\mathrm{or}~~ \bbar{\psi}_M = \psi^T_MC_M.
\end{equation}
Note that the degrees of freedom match, as we have moved from a Weyl
spinor in Euclidean space to a Majorana spinor (or equivalently a
single Weyl spinor with a (Majorana) mass term) in Lorentzian space,
each with two complex degrees of freedom off shell. In Minkowski space
the Majorana spinors take the following form in terms of 2-component
spinors (either a single left- or right-handed spinor),
\begin{equation}
  \label{eq:2compm}
  \psi_{M(-)} = \left(
    \begin{matrix}
      \xi_-\\
      \xi_-^\dagger
    \end{matrix}
  \right), \qquad
  \psi_{M(+)} = \left(
    \begin{matrix}
      \xi_+^\dagger\\
      \xi_+
    \end{matrix}
  \right),
\end{equation}
again with the caveat that one should be careful in mixing the 2-
and 4-component languages between the Euclidean and Lorentzian
theories.

\section{The Standard Model}
\label{sec:standard-model}

We have all the ingredients we need to construct the SM in flat
spacetime from an originally $SO(4)$ Euclidean theory. The SM contains
the gauge field strength terms for each group, kinetic terms for each
matter field, and Yukawa terms coupling the Higgs field to the matter
fields to give mass terms from the Higgs mechanism. A key structure is
that the weak gauge group, $SU(2)_L$, acts only on left-handed
fields. It is this chiral structure of the weak force which requires
the Yukawa interactions with the Higgs field (or some other mechanism
entirely) in order for the fermions to have mass.

We have already seen how to construct kinetic terms (and gauge field
strengths) which flow from the Euclidean theory to the proper terms
with a Lorentzian signature, for all of the fields we need. Let us now
consider the necessary Yukawa interaction terms between the Higgs and
fermion matter fields. These terms do not change form as the
background metric changes, and we can use the usual terms in the
$SO(4)$ theory.

As we must treat left- and right-handed fields differently under the
weak force, we rely on the Weyl spinors (or projections) we
constructed earlier. A common simplification is to write the SM
Lagrangian purely in terms of left-handed fields. In this form, the
right-handed fields which do not couple to the weak force are written
as antifermions of a new species of left-handed fermions. For
instance, for the up and down quarks, the left-handed $SU(2)_L$
doublet is $Q$, and the right-handed $SU(2)_L$ singlets are
$\bar{u}_R$ and $\bar{d}_R$ with the bar purely part of the name. We
then use their left-handed antiparticles, $\bar{u}, \bar{d}$, in
writing a Lagrangian.

Yukawa terms in the Euclidean theory then look just like in the
SM. For example, for the first generation of quarks (with $H$ the
Higgs scalar $SU(2)_L$ doublet),
\begin{equation}
  \label{eq:yukawae}
  Q_EH_E\bar{d}_E + Q_E\epsilon H^\dag\bar{u}_E + (\mathrm{h.c.}),
\end{equation}
where $Q_E, H_E, \bar{u}_E,$ and $\bar{d}_E$ are all Euclidean Weyl
spinors with $\gamma^5_E$ eigenvalue $-1$ and all indices are
suppressed (the $\epsilon$ tensor combines $Q_E$ and $H_E^\dag$
antisymmetrically in $SU(2)_L$ indices). Once we go to the Lorentzian
theory, the Euclidean Weyl spinors become the left-handed projections
of the SM fields, and we have exactly the SM. The leptons and other
generations all follow in the same way.

One thing to note is how the right-handed terms are generated in terms
of these left-handed fields. In the Lorentzian theory, the conjugate
of a left-handed field is right-handed, and vice versa. We do
\emph{not} have this group structure in the Euclidean theory. Thus,
when we write the Hermitian conjugate terms in the $SO(4)$ theory,
they are still fields with $\gamma^5_E$ eigenvalue $-1$. Once we are
in the Lorentzian theory, however, Weyl spinors are not
self-conjugate, and the Hermitian conjugate terms are
right-handed. After the Higgs mechanism the fermions are all (except
for the neutrino) paired up into Dirac mass terms, which mix the left-
and right-handed components.

\section{Constraints in QED}
\label{sec:constraints-qed}

As was remarked in Ref.~\cite{Mukohyama:2013ew,*emergentessay}, there is a tuning
necessary in the couplings to the clock field to reach the standard
Lorentzian theory. In this section we will restrict ourselves to the
QED sector and examine the constraints on these terms by using the
work summarized in Refs.~\cite{Kostelecky:2008ts,Fittante:2012ua} (see
references therein for details on the parameterization of operators
and the relevant experimental results). We will work in flat
(Minkowski) space with a single fermion flavor (the
electron/positron); some constraints may change in more general
settings.

We will make a connection from our model to the parameterization of
Lorentz violating operators in the Standard Model Extension (SME) used
in Refs.~\cite{Kostelecky:2008ts,Fittante:2012ua}. The SME
encapsulates the minimal set of dimension 3 and 4 CPT and Lorentz
violating operators, and constructs observables which can be
constrained by experiment. The general Minkowski space QED Lagrangian
(with electromagnetic tensor $F_{\mu\nu}$ and fermion $\psi$) in the SME is
\begin{equation}
  \label{eq:sme}
  \mathcal{L} = \frac{i}{2}\bbar{\psi}\Gamma_\nu\stackrel{\leftrightarrow}{\partial^\nu}\psi - \bbar{\psi}
M\psi - \frac{1}{4}K_{\mu\nu} F^{\mu\nu},
\end{equation}
with
\begin{align}
  \label{eq:smeGM}
  \Gamma_\nu &\equiv \gamma_\nu + c_{\mu\nu}\gamma^\mu + d_{\mu\nu}\gamma^5\gamma^\mu + e_\nu \nonumber\\
      &~~+ if_\nu\gamma^5 + \half g_{\lambda\mu\nu}\Sigma^{\lambda\mu},\\
  M &\equiv m + a_\mu\gamma^\mu + b_\mu\gamma^5\gamma^\mu + \half H_{\mu\nu}\Sigma^{\mu\nu},\\
  K_{\mu\nu} &\equiv  F_{\mu\nu} -2\left(k_{AF}\right)^\kappa\epsilon_{\kappa\lambda\mu\nu}A^\lambda + \left(k_F\right)_{\kappa\lambda\mu\nu}F^{\kappa\lambda},\label{eq:smeK}
\end{align}
where $\half\Sigma^{\mu\nu} \equiv \frac{i}{4}[\gamma^\mu, \psi^\nu]$ and all $\gamma^\mu$ are in
Minkowski space. The observables are combinations of the free
parameters $a_\mu, b_\mu, c_{\mu\nu}, d_{\mu\nu}, e_\nu, f_\nu,
g_{\lambda\mu\nu}, H_{\mu\nu}, (k_{AF})^\kappa,$ and
$(k_F)_{\kappa\lambda\mu\nu}$ (see Refs.~\cite{Kostelecky:2008ts,
  Fittante:2012ua} for the precise definitions and counting of
independent parameters and observable combinations).

Let us start with the photon. We can parameterize any deviation from
the interaction term with the clock field which leads to the
Lorentzian theory as
\begin{equation}
  \label{eq:photonlve}
  \frac{4}{M^4}(1 + \epsilon_A)F^{\mu\rho}_E F^\nu_{E\rho}\partial_\mu\phi\partial_\nu\phi,
\end{equation}
with $\epsilon_A$ the deviation from Eq.~\eqref{eq:vector_e}. In the
Lorentzian theory then, we end up with the additional term
\begin{equation}
  \label{eq:photonlvm}
  \epsilon_A \delta_{ij} F^{0i} F^{0j}.
\end{equation}
This is a CPT even operator, which violates Lorentz invariance,
corresponding to the SME parameter $(k_F)_{\kappa\lambda\mu\nu}$ in
Eq.~\eqref{eq:smeK}: it is constrained to have $|\epsilon_A| <
\mathcal{O}(10^{-32})$ (cf.~the observables $\tilde{\kappa}$, in
particular the component $\tilde{\kappa}_{e+}^{ZZ}$ in
Ref.~\cite{Kostelecky:2008ts}).

In the matter sector, we parametrize a deviation from
Eq.~\eqref{eq:dirace}, which gives the proper Minkowski Lagrangian in
$\mathcal{M}_0$, with the parameters $\epsilon_{\psi_{1,2}}$ as
\begin{align}
  \label{eq:electronlve1}
  \frac{1}{2M^2}\delta^{\mu\nu} &\left[ (1 + \epsilon_{\psi_1}) \left(i\bbar{\psi}_E\gamma^5_E\stackrel{\leftrightarrow}{\partial_\mu}\psi\right) \right.\nonumber\\
&\left. - (1 + \epsilon_{\psi_2})\left(i\bbar{\psi}\gamma^\rho_E\stackrel{\leftrightarrow}{\partial_\mu}\psi\right)\partial_\rho \phi\right]\partial_\nu\phi.
\end{align}
We then have the following Lorentz violating operators in the theory
in Minkowski space, the first of which is CPT even, the second CPT
odd:
\begin{equation}
  \label{eq:electronlve2}
  \frac{i}{2}\epsilon_{\psi_1}\left(\bbar{\psi}\gamma^0_M\stackrel{\leftrightarrow}{\partial_0}\psi\right)
+ \half \epsilon_{\psi_2}\left(\bbar{\psi}\gamma^5_M\stackrel{\leftrightarrow}{\partial_0}\psi\right).
\end{equation}
However, through a field redefinition this second operator ($f_0$ in
the SME above) can actually be removed at leading order (in
$\epsilon_{\psi_2}$) and absorbed into $\epsilon_{\psi_1}$ at second
order (see the discussion in
Refs.~\cite{Kostelecky:2008ts,Fittante:2012ua} and references
therein). Thus we do not have CPT violation, regardless of the precise
tuning, contrary to what was stated originally in
\cite{Mukohyama:2013ew,*emergentessay}. The CPT even operator must
have coefficient $|\epsilon_{\psi_1}| < \mathcal{O}(10^{-15})$
(corresponding to $\tilde{c}_{TT}/m_e$ in
Ref.~\cite{Kostelecky:2008ts}) and this gives a constraint, through
field redefinition, of $|\epsilon_{\psi_2}| < \mathcal{O}(10^{-7})$.

We have seen that there is a precise tuning in the couplings of the SM
fields to the clock field needed to avoid Lorentz violation
constraints. Besides the tuning in these coefficients, there are other
possible terms which can be dangerous, as noted in
Ref.~\cite{Mukohyama:2013ew,*emergentessay}. Of the 10 terms which are scalars under
$SO(4)$, Hermitian, and include at most one derivative acting on
spinors, we have the usual mass and kinetic terms, and the 2 terms we
have already included. There are 4 additional terms with couplings to
the clock field,
\begin{align}
  \label{eq:addterms}
  \left(\bbar{\psi}\gamma^\mu_E\psi\right)\partial_\mu\phi,& \quad \left(i\bbar{\psi}\gamma^5_E\gamma^\mu_E\psi\right)\partial_\mu\phi,\\
  \delta^{\mu\nu}\left(i\bbar{\psi}\stackrel{\leftrightarrow}{\partial_\mu}\psi\right)\partial_\nu\phi,& \quad \delta^{\mu\nu}\left(i\bbar{\psi}\gamma^5_E\gamma^\rho_E\stackrel{\leftrightarrow}{\partial_\mu}\psi\right)\partial_\rho\phi\partial_\nu\phi. \nonumber
\end{align}

The first term corresponds, in the Lorentzian theory, to a $\gamma^5$ mass
term, which can be removed through a chiral transformation. The third
term (corresponding to $e_\mu$ in the SME) is CPT and Lorentz
violating, but can also be removed by transformations and field
redefinitions (it can be absorbed into $a_\mu$ and is not observable
with a single flavor in flat space; see the summary in
Ref.~\cite{Fittante:2012ua} and references therein). The second term
($b_T$ in the SME) is CPT odd and Lorentz violating, constrained to be
less than $\mathcal{O}(10^{-27}~\mathrm{GeV})$ (see the combinations
$\tilde{b}_T$ and $\tilde{g}_T$ in
Ref.~\cite{Kostelecky:2008ts}). This is problematic as the generated
mass scale is presumably $\sim M$, and there does not appear to be a
simple way to forbid such a term. Finally, the fourth term, which is
CPT even and Lorentz violating, is constrained to be less than
$\mathcal{O}(10^{-24})$ (constrained via the tracelessness of
$d_{\mu\nu}$; see the observable $\tilde{d}_+$ in
Ref.~\cite{Kostelecky:2008ts}). Again, there is not an obvious way to
forbid such a term, and its pure number coefficient is a free
parameter.

Finally, we also have two terms which do not involve interactions with
the clock field. One is the standard $\gamma^5_E$ mass term,
$\bbar{\psi}\gamma^5_E\psi$, which we will transform away in the
Euclidean theory (or it corresponds to the unobservable parameter
$a_0$ in the SME). The second term is
$\bbar{\psi}\gamma^5_E\gamma^\mu_E\stackrel{\leftrightarrow}{\partial_\mu}\psi$,
which we can write in the Lorentzian theory as
\begin{align}
  \label{eq:extraterm}
  &\bbar{\psi}\left(i\gamma^0_M\gamma^5_M\stackrel{\leftrightarrow}{\partial_0} + \gamma^0_M\gamma^i_M\stackrel{\leftrightarrow}{\partial_i}\right)\psi \nonumber\\
  &= -i\bbar{\psi}\left(\gamma^5_M\gamma^0_M\stackrel{\leftrightarrow}{\partial_0} + \Sigma^{0i}\stackrel{\leftrightarrow}{\partial_i}\right)\psi,
\end{align}
where we used that $\gamma^0_M\gamma^i_M = \half[\gamma^0_M,\gamma^i_M] +
\half\{\gamma^0_M,\gamma^i_M\} = \half[\gamma^0_M,\gamma^i_M]$. The
first term in Eq.~\eqref{eq:extraterm} is the same as the last term
discussed in the previous paragraph, and thus has the same
constraint. In the SME, the second term is a component of the trace
part of the parameter $g_{\mu\nu\lambda}$ (the coefficient of a CPT
odd and Lorentz violating operator), $g^{(T)}_\mu \equiv
g_{\mu\nu}^{~~~\nu}$ (note that $g_{000}$ does not contribute since
$\Sigma^{00} = 0$). This is not an observable component of $g$ as it
can be removed through a field redefinition (see
Ref.~\cite{Fittante:2012ua} and references therein).

\section{Discussion, Conclusion, and Outlook}
\label{sec:disc-concl-outl}

This article follows the idea that the apparent Lorentzian dynamics of
usual field theories is an emergent property and that the underlying
field theory is in fact strictly Riemannian. This requires the
introduction of the clock field, a scalar field playing the role of
the physical time. The microscopic theory is Euclidean, and time
evolution is just an effective and emergent property, which holds on
some energy scales, and in some regions of the Euclidean
space. Through interactions with the clock field the effective theory
flows to the standard Lorentzian picture.

In Ref.~\cite{Mukohyama:2013ew,*emergentessay}, we were able to
perform a construction in flat spacetime for scalar, vector, and Dirac
spinors restricted to classical fields. In order for all the fields to
propagate in the same emergent Lorentzian metric, the couplings to the
clock field needed to be adjusted with care. This work was a proof of
concept in constructing a model with the Lorentzian metric only
emerging at energies below the vev of the gradient of the clock field,
with many open and interesting questions. In this work we have
addressed several of these questions.

The present analysis has shown that it is possible to construct a
Euclidean theory with fermions that reduce, once the gradient of the
clock field has a vev on ${\cal M}_0$, to Lorentzian Weyl and Majorana
fermions. This completes the basic fields needed in the Standard Model
and common extensions. The clock field allows us to avoid the typical
difficulties in constructing Euclidean theories of these types of
fermions. We then showed that it is possible to construct a Euclidean
theory leading to an emergent version of the Standard Model by adding
the Standard Model structure to the dynamics necessary for the
emergence of a Lorentzian metric.

To finish, we have analyzed with care the fine-tuning required to
ensure CPT and Lorentz invariance. One crucial point is that the terms
necessary in our model do not induce CPT violation. Bounds on the
deviations from the adjusted couplings to the clock field, as well as
other possible interaction terms in this framework, can be obtained
from experimental QED constraints. Forbidding additional operators and
ensuring the value of the necessary coupling constants is an open
question.

There are still many interesting future directions to pursue in this
framework for emergent Lorentz symmetry. One would like to move beyond
the classical level and quantize the theory, as well as understand the
mechanism which leads to the vev of the clock field. The possible
violation of CPT and Lorentz symmetry also needs to be investigated
further. Even with these and other open questions, we now have a basic
model which can reproduce the Standard Model and its Lorentzian
background with time evolution from a purely Riemannian theory with no
concept of time.

\begin{acknowledgments}
  \noindent
  This work was supported by the World Premier International Research
  Center Initiative (WPI Initiative), MEXT, Japan. This work was made
  in the ILP LABEX (under reference ANR-10-LABX-63) and was supported
  by French state funds managed by the ANR within the Investissements
  d'Avenir programme under reference ANR-11-IDEX-0004-02 and by the
  ANR VACOUL, ANR-10-BLAN-0510. One of us (SM) acknowledges the support
  by Grant-in-Aid for Scientific Research 24540256 and 21111006. 
\end{acknowledgments}

\appendix*
\section{Gamma Matrices and Fermions}
\label{apx:gamma-matr-ferm}

In this appendix we will try to motivate some of the choices made in
our Euclidean formulation of fermions. Our procedure will be to take
the action proposed in the Euclidean theory as an ansatz, and require
that we end with a proper Lorentzian theory. This will then define the
relationship between the representations of the gamma matrices (which
will not be chosen a priori) and identifications between quantities in
the two theories.

Let us start with a massless Dirac fermion in the $SO(4)$ theory,
coupled to the clock field as in the action of Eq.~\eqref{eq:dirace},
\begin{align}
  \label{eq:app_dirace}
  S_\psi &= \int\mathrm{d}^4x\left\{\bbar{\psi}_E\left( \frac{i}{2}\gamma^\mu_E\stackrel{\leftrightarrow}{\partial_\mu} - m \right)\psi_E \right.\\
      &\left. + \frac{1}{2M^2}\delta^{\mu\nu}\left[ \left(i\bbar{\psi}_E\gamma^5_E\stackrel{\leftrightarrow}{\partial_\mu}\psi\right) -\left(i\bbar{\psi}\gamma^\rho_E\stackrel{\leftrightarrow}{\partial_\mu}\psi\right)\partial_\rho \phi\right]\right\},\nonumber
\end{align}
but without assuming the form of $\bbar{\psi}_E$ or $\gamma^\mu_E$.

Although we can form an $SO(4)$ invariant with $\psi_E^\dag\psi_E$, we wish to
mirror the usual Lorentzian construction, so we have used
$\bbar{\psi}$. In order for this to transform as $\psi_E^\dag$
(i.e.~in the opposite way of $\psi_E$), any matrix we attach to
$\psi_E^\dag$ to form $\bbar{\psi}_E$ must commute with the $SO(4)$
generators. Thus we have
\begin{equation}
  \label{eq:app_psibare}
  \bbar{\psi}_E \equiv \psi_E^\dag\gamma^5_E.
\end{equation}

After the clock field's derivative has a vev $M^2$, chosen to define
the $t$-direction, the action becomes
\begin{equation}
  \label{eq:2}
  \mathcal{S}_\psi \rightarrow \int\mathrm{d}t\mathrm{d}^3x~\left(\bbar{\psi}i\gamma^i_E\partial_i\psi\right) + i\left(\bbar{\psi}\gamma^5_E\partial_0\psi\right).
\end{equation}
Since the clock field has now picked out a direction, morphing $SO(4)$
to $SO(3,1)$, we expect that we should now have a free fermion
propagating in Minkowski space. We recover the usual action,
\begin{equation}
  \label{eq:10}
  \mathcal{S}_M = \int\mathrm{d}t\mathrm{d}^3x~i\bbar{\psi}\gamma^\mu_M\partial_\mu\psi = i\bbar{\psi}\left(\gamma^0_M\partial_0 + \gamma^i_M\partial_i\right)\psi
\end{equation}
by identifying\footnote{Note that usually $\beta$ and $\gamma^0$ are
  used interchangeably because they are numerically the same. However,
  at least in the chiral representation, the spin structure is
  different.}
\begin{align}
  \label{eq:3}
  \bbar{\psi} \equiv \psi^\dagger\gamma^5_E &~~ \rightarrow ~~\bbar{\psi}_M \equiv \psi^\dagger_M\beta, \nonumber\\
  \gamma^i_E \rightarrow \gamma^i_M, &\qquad \gamma^5_E \rightarrow \gamma^0_M.
\end{align}

From these identifications and the definition of $\gamma^5_E$ we know that
$\{\gamma^5_E, \gamma^i_M\} = 0$ and $[\gamma^5_E, \gamma^0_M] =
0$. Therefore $\gamma^5_E\gamma^\mu_M(\gamma^5_E)^{-1} =
(\gamma^\mu_M)^\dagger$, with $\gamma^0_M$ Hermitian and $\gamma^i_M$
anti-Hermitian (from the definition of the Clifford algebra as
$\{\gamma^\mu_M, \gamma^\nu_M\} = -2\eta^{\mu\nu}$). Combined with
$\gamma^5_E$ being Hermitian, or by direct computation, we find that
it has the right properties (see, e.g., Appendix G of
Ref.~\cite{Dreiner:2008tw}) to be the matrix $\beta$: $\bbar{\psi}_M$
transforms oppositely of $\psi_M$ such that $\bbar{\psi}_M\psi_M$ is a
(Hermitian) Lorentz scalar. Furthermore, the Clifford algebra for
$SO(4)$ tells us that $(\gamma^0_E)^2 = -1$ and we chose an
anti-Hermitian representation, $(\gamma^0_E)^\dagger = -\gamma^0_E$,
such that the $SO(4)$ generators we defined were Hermitian. Since
$\gamma^0_E$ anticommutes with all the $\gamma^\mu_M$ this implies
\begin{equation}
  \label{eq:4}
  \gamma^0_E = i\gamma^5_M.
\end{equation}
All the above requirements are then consistent, coming from the
proposed $SO(4)$ action. The gamma matrices all match what was given
in Sec.~\ref{sec:diracferm}.

For Weyl spinors (Euclidean spinor eigenstates of $\gamma^5_E$) we can
follow the same procedure, and we find that we reach the Minkowski
Weyl action with the same identification of gamma matrices, including
that the spinor is now an eigenstate of $\gamma^5_M$.


\nocite{apsrev41Control}
\bibliographystyle{apsrev4-1}
\bibliography{sm_cons_refs}

\end{document}